# Novel Single Clad Ho-doped Fiber with High Slope Efficiency and Low Ion Pairing

Robert E. Tench, Senior Member, *IEEE*, Wiktor Walasik, Alexandre Amavigan,
Jean-Marc Delavaux, Senior Member, *IEEE*, Colin C. Baker, and Daniel Rhonehouse

*Abstract*— We report the design and experimental and simulated performance for a 2050 nm band fiber amplifier with high optical-optical slope efficiency and low ion pairing, using a novel high performance single clad Ho-doped fiber from the Naval Research Laboratory (NRL). We measure an optical-optical slope efficiency of 57% using 1 mW input signal power and 1860 nm pumping which we believe is the highest slope efficiency obtained to date for a single clad single stage copumped HDFA. A new method for non-destructive measurement of the ion pairing coefficient in Ho-doped fibers is introduced and validated. Using this method, we link our 57% slope efficiency to a low ion pairing coefficient of 4% in the NRL Ho-doped fiber as derived from our experimental data. We present an overview and survey of the ion pairing results for Ho-doped fiber amplifiers and lasers reported so far in the literature.

## I. INTRODUCTION

Current progress in advanced infrared fiber amplifiers [1-8], high power sources for space applications (including earth-to-satellite and satellite-to-satellite communications) [9, 10], and high power single frequency sources for advanced gravity wave detection [11, 12] highlights the need for large spectral bandwidth, high efficiency Ho-doped fiber optical amplifiers (HDFAs) in the eye safe region from 2000 nm—2150 nm. Both PM (polarization maintaining) and non-PM HDFAs exhibiting high optical-optical power conversion efficiency with low or moderate signal input powers are particularly attractive for SWAP (size, weight, and power) optimization in many emerging applications.

In this paper we present the design and experimental results for a 2050 nm band fiber amplifier with high optical-optical slope efficiency and low ion pairing, using a novel high performance single clad Ho-doped fiber from the Naval Research Laboratory (NRL). Here we report a measured optical-optical slope efficiency of 57% using 1 mW input signal power which we believe is the highest slope efficiency measured to date for a single clad single stage in-band-copumped HDFA with low signal input powers. This efficiency is linked to a low ion pairing coefficient of 4% in the doped fiber derived from our data, using a novel non-destructive method of determining the ion pairing coefficient.

Our paper is organized as follows. Section II contains a survey of recent work that is focused on the effects of pair-induced quenching (PIQ) or equivalently ion pairing in Ho-doped fibers. Section III describes the characteristics and specifications of the novel Ho-doped fiber from NRL. Section IV presents the experimental setup, our approach to simulations, and outlines the new non-destructive method of accurately and precisely determining the ion pairing coefficient for in-band pumped rare-earth-doped fiber amplifiers and lasers. Section V contains experimental data and analysis leading to an accurate determination of the ion pairing coefficient in the NRL fiber. Following discussions in Section VI, the paper concludes with a summary in Section VII.

## II. SURVEY OF RECENT WORK ON PIQ/ION PAIRING IN HDFAs

To begin the survey, reference [9] from 2020 reports the realization and simulation of high-power holmium doped fiber lasers for long-range transmission. A cutback method is employed to measure the ion pairing in the Ho doped fiber. For Exail PM Ho-doped fiber IXF-HDF-PM-8-125, the ion pairing is found to be 15 ± 1%.

Reference [13] from 2025 presents a study of Ho-doped silica fiber lasers combining high doping and high efficiency at 2100 nm operating wavelength. Both core-pumped and pedestal cladding-pumped fibers are investigated. The authors measure an output power of 22.5 W and a slope efficiency of 81% with a 1940 nm pump source. This high level of performance is consistent with a favorable Ho-doped fiber composition with relatively low ion pairing of 10%.

Reference [14] from 2024 reports a detailed study of the simulation and experimental performance of Exail Ho-doped

Robert E. Tench (robert.tench@retandassociatesllc.com) is with RET and Associates LLC, 6081 Hamilton Boulevard, Suite 600-606, Allentown, PA 18106. At the time of this work he was with Cybel LLC.

Wiktor Walasik (wiktor.walasik@caci.com) is with CACI, 15 Vreeland Road, Florham Park, NJ 07932. At the time of this work he was with Cybel LLC.

Alexandre Amavigan (alexandre.amavigan@cybel-llc.com) and Jean-Marc Delavaux (jm@cybel-llc.com) are with Cybel LLC, 62 Highland Drive, Bethlehem, PA 18017.

Colin C. Baker (colin.c.baker.civ@us.navy.mil) and Daniel Rhonehouse (daniel.l.rhonehouse.civ@us.navy.mil) are with the Naval Research Laboratory, 4555 Overlook Ave., SW, Washington, DC 20375.

Color versions of one or more of the figures in this article are available online at http://ieeexplore.ieee.org

fiber IXF-HDF-PM-8-125. A newly invented method of pumping the HDFA using broadband ASE is introduced and validated. The ion pairing coefficient of the Exail PM Ho-doped fiber is measured to be 13.5 ± 1 % in this work.

Reference [15] from 2025 demonstrates and analyzes the performance of a continuous all-fiber Ho laser delivering up to 180 W at 2.12 μm in a single spatial mode, with an optimum optical efficiency of 60 %, based on a triple clad holmium doped fiber. The gain fiber is Exail 3CF-Ho-20-105. While no direct measurements of ion pairing/PIQ are reported here, the high level of laser performance indicates that the fiber likely has a favorably low level of ion pairing.

Reference [16] from 2025 is a detailed report of the design and performance of a novel compact 2121nm 2W PM Ho-doped fiber amplifier for molecular hydrogen detection. Successful high power operation at the quite high wavelength of 2121 nm is achieved in spite of the 13.5 ± 1% ion pairing in the Exail IXF-HDF-PM-8-125 Ho-doped fiber.

Reference [17] from 2025 is a study of pair-induced quenching in holmium doped fiber lasers and its impact on efficiency and operating wavelength. This paper outlines that while the low-concentration Ho doped fibers intended for core-pumped applications have reached ion pairing coefficients in the high single percentage points around 10%, thus managing above 85 % efficiency, the highly doped fibers for clad-pumped applications used in experiments have shown ion pairing of around 30 %, although recent improvements suggest that newer fibers have substantially improved upon this.

Reference [18] from 2025 outlines the effect of ion pair dynamics on unsaturable absorption in holmium doped fibers. Here a Ho-doped fiber manufactured by the Czech Academy of Sciences (IPE) designated as NP1558 is determined to have an ion pairing coefficient of 21.2%.

Reference [19] from 2025 describes the design and experimental performance of enhanced long-wavelength emission fibers and 2109 nm lasing in Tm3+/Ho3+ co-doped germanate-core/silicate-cladding glass fiber. Using a 2-cm-long fiber as the gain medium, an all-fiber laser emitting at 2109 nm was demonstrated, exceeding the wavelength of reported Tm3+/Ho3+ co-doped fiber lasers. While no direct measurements of ion pairing in the Ho host were reported, the favorable long wavelength performance of this novel design indicates that a low degree of ion pairing is likely.

Reference [20] from 2025 presents the design and performance of an efficient 2 μm polarization-maintaining Ho-doped fiber laser with intracavity core pumping. In the simulation work, a rate equation model for the intracavity Tm and Ho laser is established to provide theoretical guidance for the experiment. In the experimental work, a maximum output power of 12.1 W/11.8 W is realized with an optical-to-optical efficiency of 37.8%/36.9% and a slope efficiency of 39.2%/39.1% from 792 nm to 2050.7 nm/2088.7 nm. In the Ho laser, a piece of 1-m-long PM HDF is used as the gain fiber of the internal cavity. The gain fiber has a core/cladding diameter of 10/130 μm and the corresponding NA of 0.146/0.46. The experimental output powers at 2050 nm are about 5% less than the simulated values, indicating that the ion pairing in the fiber is likely to be in the range of mid- to high single digits in percentage.

Reference [21] from 2025 reports new results on a low threshold, high-power Tm/Ho co-doped double-clad single-mode silica fiber operation at 2.08 μm. The Tm-to-Ho concentration ratio in the fiber was 10:1. A measured slope efficiency of 45% relative to the pump power indicates that the ion pairing in the Ho host may be somewhat larger than mid- to high single digits in percentage, but this conclusion needs to be verified through further detailed studies.

Reference [22] is a 2024 conference publication that presents a brief summary of some of the work outlined in detail in the present manuscript.

Finally, reference [23] from 2019 is a comprehensive study of the effects of ion clustering and excited state absorption on the performance of Ho-doped fiber lasers. This work presents a detailed theoretical model and analysis of the processes involved with ion pairing/PIQ in Ho-doped fibers and then outlines an experimental study of a Ho-doped fiber from Nufern. The ion pairing coefficient for Nufern SM-HDF-10-130 Ho-doped fiber is found to be 10%.

### III. NOVEL HO-DOPED FIBER FROM NRL

The NRL Ho-doped step-index single clad silica fiber in our experiments has a core diameter of 10 μm, a cladding diameter of 92 μm, a core-cladding refractive index difference of 1.2 × $10^{-2}$, and a numerical aperture of NA = 0.186. These parameters are illustrated in the refractive index profile shown in Figure 1. This profile was measured with an IFA-100 RIP system.

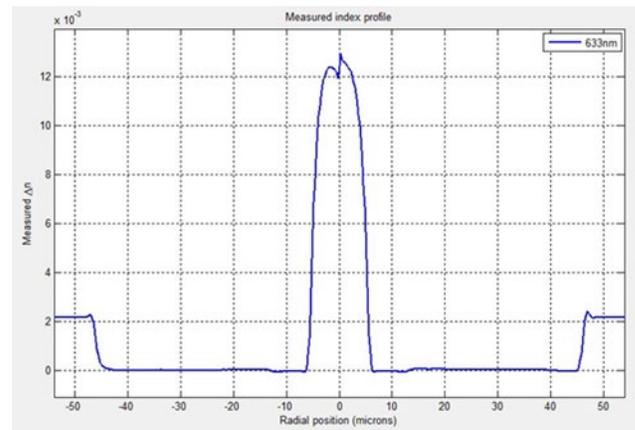

Figure 1. Refractive index profile of the NRL Ho-doped fiber.

The Ho ion concentration in the core is 0.7%-wt, and the peak absorption coefficient is 51 dB/m at 1940 nm. The Al co-doping of the core was selected for maximum power conversion efficiency with low input signals. Figure 2 shows the measured absorption coefficient of the fiber as a function of wavelength from 1700 nm to 2200 nm.

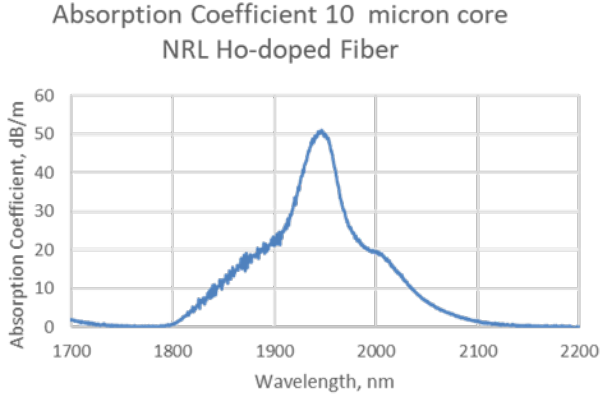

Figure 2. Experimental absorption coefficient of the NRL Ho-doped fiber.

## IV. EXPERIMENTAL SETUP AND APPROACH TO SIMULATIONS

We measured the performance of the NRL fiber in the single stage copumped amplifier architecture shown in Figure 3. Here the Ho-doped fiber is F1, isolators I1 and I2 in the signal path prevent feedback from external reflections and establish unidirectional operation, the pump source is P1, and the wavelength division multiplexer WDM1 couples the input signal and the pump source into F1 with low loss.

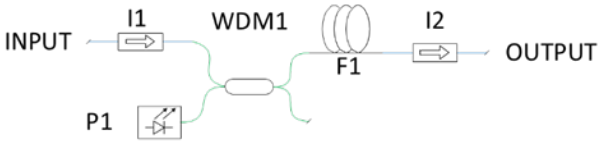

Figure 3. Experimental setup for measuring the performance of the Ho-doped fiber in a single stage amplifier.

In our measurements, we employed in-band Tm-doped fiber laser pump sources at both 1860 nm and 1940 nm as outlined in [7, 8, 14, 22]. Typical pump powers ranged from 0 to 2.5 W CW.

Initial simulations for the HDFA in Figure 3 were carried out using the methods, rate equations, and number density equations presented in [7, 14]. We employed McCumber theory to calculate the gain coefficient as a function of wavelength using the measured absorption coefficient vs. wavelength shown in Figure 2 [24].

The length of the NRL Ho-doped fiber in our experimental setup was determined by initial simulation studies where the pump wavelength of P1 was set to 1860 nm, the pump power ranged from 0—2.5 W, and the input signal was 1 mW (0 dBm) at a wavelength of 2050 nm. Signal powers and pump powers were simulated at the input and output of the active Ho-doped fiber F1. With these pump and signal parameters, and assuming a range of ion pairing coefficients from 0—15%, we determined a simulated optimum length of the NRL fiber for maximum signal output power of 2.2—2.5 m. We therefore chose a fiber length of 2.5 m for our experiments going forward.

Our approach to the next analysis and simulations was guided by our earlier studies using a polarization-maintaining Ho-doped fiber from Exail, IXF-HDF-PM-8-125 [1, 2, 7, 8, 14]. This approach is illustrated by the plots shown in Figure 4 for the performance of the single stage HDFA of Figure 3 using the Exail Ho-doped fiber. Here the pump power was fixed at 2.47 W.

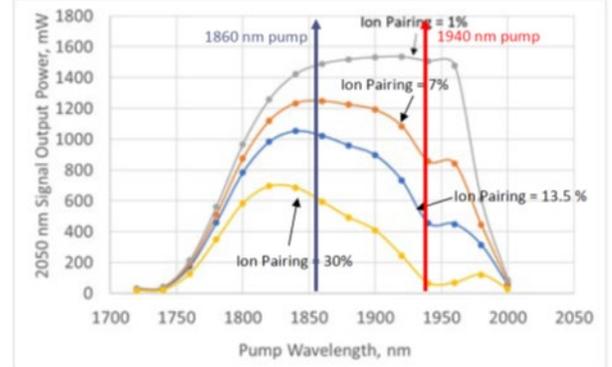

Figure 4. Simulated output signal power of single stage HDFA using Exail IXF-HDF-PM-8-125, as a function of pump wavelength and ion pairing coefficient. The vertical arrows show the key pump wavelengths of 1860 nm (in blue) and 1940 nm (in red).

In the simulations of Figure 4, the total co-pump power is 2.47 W, the length of the Exail fiber is 2.0 meters, and the input signal is 1 mW at 2051 nm. The x-axis of the plot shows pump wavelengths from 1700 nm to 2050 nm, and the y-axis is amplified signal output power from 0 to 1800 mW. The parameter in the plot is the degree of ion pairing assumed in the Exail Ho-doped fiber.

We note that as reported in [1, 2, 7, 8, 14] our simulations agree quite well with the experimental data for this Exail fiber and so predict its performance accurately and precisely.

It is apparent from Figure 4 that as the ion pairing coefficient varies from its minimum of 1% to a maximum of 30%, the simulated signal output power as a function of pump wavelength depends strongly and in a completely predictable manner on the degree of ion pairing in the fiber. We therefore observe that by measuring the ratio of signal output powers for the 1860 nm and 1940 nm pump wavelengths (shown with vertical arrows in Figure 4), the degree of ion pairing in the in-band-pumped Ho-doped fiber can be accurately determined in a non-destructive manner. This new method is the basis for our experiments and analysis going forward in the next section.

## V. DETERMINATION OF ION PAIRING COEFFICIENT FOR THE NRL HO-DOPED FIBER

Following the simulations and analysis of the previous section, we then measured the performance of a 2.5 meter length of the NRL Ho-doped fiber using the experimental setup in Figure 3. These measurements are presented in Figure 5 where the x-axis is pump power from 0 to 2.5 W and the y-axis is 2051 nm amplifier signal output power from 0 to 1000 mW. The signal input power for these measurements was 1 mW (0 dBm).

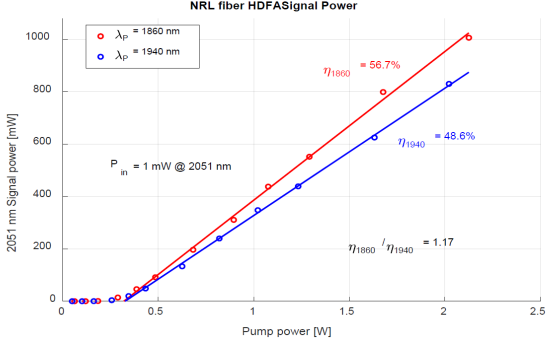

Figure 5. Signal output power for the NRL Ho-doped fiber amplifier vs. pump power for 1860 nm and 1940 nm pump wavelengths.

From these data, we observe that the slope efficiency for 1860 nm pumping is 57%, while the slope efficiency for 1940 nm pumping is considerably lower at 49%. We also observe that at a pump power of 1.3 W, the amplified signal output powers from the HDFA are 552 mW for 1860 nm pumping and 470 mW for 1940 nm pumping.

With these data in mind, we then simulated the output power for the HDFA for 1.3 W pump powers at 1860 nm and 1940 nm as a function of the ion pairing coefficient. The results of these simulations are presented in Figure 6 where the x-axis is ion pairing from 0 % to 16% and the y-axis is signal output power from 0 to 800 mW. For a ratio of 552 mW/470 mW = 1.17, we then see that the ion pairing for the NRL Ho-doped fiber is 4%.

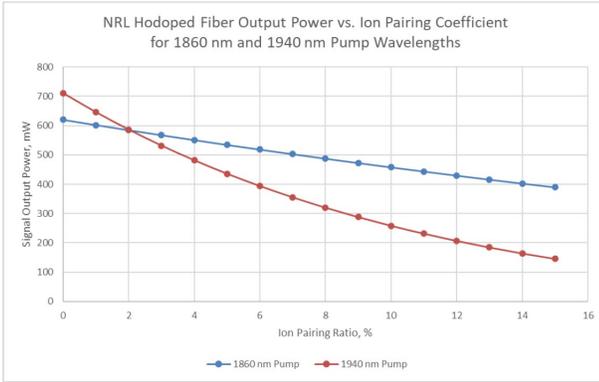

Figure 6. Signal output power from the NRL HDFA as a function of ion pairing coefficient from 0% to 16%. The pump power is 1.3 W.

The typical relationship between the ion pairing coefficient and the ratio of 1860 nm/1940 nm amplified signal output powers can also be expressed with the power ratio plotted on the x-axis and the ion pairing coefficient plotted on the y-axis as shown in Figure 7.

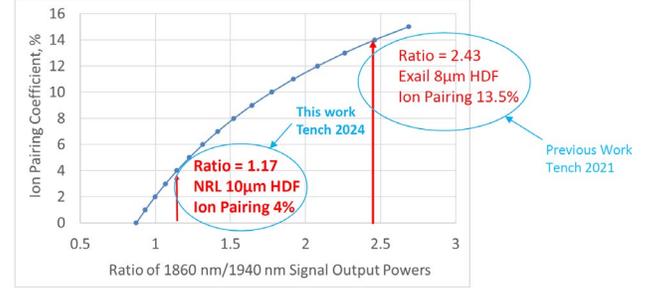

Figure 7. Ion pairing coefficient as function of the ratio of signal output powers for 1860 nm and 1940 nm.

Here we indicate both the results for the novel NRL Ho-doped fiber [22 and this work] and for the commercially available Exail Ho-doped fiber IXF-HDF-PM-8-125 [1, 2, 7, 8, 14]. The NRL Ho-doped fiber is clearly superior in its power conversion efficiency for low signal input powers.

## VI. DISCUSSIONS

Table I summarizes the results published to date for measurements and determinations of the ion pairing coefficient for Ho-doped fibers from four separate manufacturers. We see that the ion pairing values range from a high of 21.2 % [18] to a record low of 4% [22 and this work]. Two separate determinations of the ion pairing for Exail IXF-HDF-PM-8-125 using two different methods as reported in [9] and [14] agree well with one another.

| Fiber ID | Reference | Ion Pairing Coefficient |
|---|---|---|
| Czech Academy of Sciences IPE NP1558 | [18] | 21.2% |
| Exail IXF-HDF-PM-8-125 | [9] | 15 ± 1% |
| Exail IXF-HDF-PM-8-125 | [14] | 13.5 ± 1% |
| Czech Academy of Sciences IPE | [13], [17] | 10% |
| Nufern SM-HDF-10/130 | [23] | 10% |
| NRL | [22], this work | 4% |

Table 1. Measured and derived ion pairing coefficients reported in the literature.

We additionally note that the new method of determining the ion pairing coefficient using the ratio of amplified signal output powers for two selected in-band pump wavelengths is not confined only to Ho-doped fibers. This method is immediately applicable to all rare-earth-doped fiber amplifiers using in-band pumping within the manifold that includes overlapping absorption coefficients for the pump and gain coefficients for the signal. As a result, the method outlined here can be used to non-destructively measure the ion pairing coefficient for many other in-band pumped rare earth doped fiber amplifiers including Er, Yb, and Tm. This will be explored further in future publications.

## VII. SUMMARY

We have reported the experimental and simulated results for a 2050 nm band fiber amplifier with high optical-optical slope efficiency and low ion pairing, using a novel high performance single clad Ho-doped fiber from the Naval Research Laboratory (NRL). In our experiments, we measured an optical-optical slope efficiency of 57% using 1 mW input signal power and 1860 nm pumping which we believe is the highest slope efficiency obtained to date for a single clad single stage copumped HDFA.

We also presented a new method for non-destructive measurement of the ion pairing coefficient in Ho-doped fibers. Using this method, we link our 57% slope efficiency to a low ion pairing coefficient of 4% in the NRL Ho-doped fiber as derived from our experimental data. This value is the lowest ion pairing coefficient reported to date and represents a step function improvement in the performance of single clad Ho-doped fiber amplifiers.

We additionally surveyed the results published so far on the measurement and determination of ion pairing coefficients in Ho-doped fibers from several manufacturers. Our comparison of the results indicates that the novel NRL Ho-doped fiber is superior in its performance to all other Ho-doped fibers studied to date.

We believe that the advances reported here will contribute significantly to the future development, applications, and progress of Ho-doped fiber amplifiers in many important venues such as LIDAR, optical sensing (including the high sensitivity detection of trace gases), aerospace and space applications (including earth-to-satellite and satellite-to-satellite communications), and advanced gravity wave detection.